\documentclass{emulateapj}
\usepackage{apjfonts}

\newcommand{\heii}{\ion{He}{2} $\lambda 4686$}
\newcommand{\heia}{\ion{He}{1} $\lambda 5876$}
\newcommand{\heib}{\ion{He}{1} $\lambda 4471$}

\shorttitle{VELOCITY-DELAY MAPS FOR ARP~151}
\shortauthors{BENTZ ET AL.}

\begin{document}

\title{The Lick AGN Monitoring Project: Velocity-Delay Maps from the
  Maximum-Entropy Method for Arp~151}

\author{ 
Misty~C.~Bentz\altaffilmark{1,2},
Keith~Horne\altaffilmark{3},
Aaron~J.~Barth\altaffilmark{1},
Vardha~Nicola~Bennert\altaffilmark{4},
Gabriela~Canalizo\altaffilmark{5,6},
Alexei~V.~Filippenko\altaffilmark{7},
Elinor~L.~Gates\altaffilmark{8},
Matthew~A.~Malkan\altaffilmark{9},
Takeo~Minezaki\altaffilmark{10},
Tommaso~Treu\altaffilmark{4},
Jong-Hak~Woo\altaffilmark{11}, and
Jonelle~L.~Walsh\altaffilmark{1}
}

\altaffiltext{1}{Department of Physics and Astronomy,
                 4129 Frederick Reines Hall,
                 University of California,
                 Irvine, CA 92697;
                 mbentz@uci.edu .} 

\altaffiltext{2}{Hubble Fellow.}

\altaffiltext{3}{SUPA Physics and Astronomy, 
                 University of St Andrews, 
                 North Haugh, St Andrews KY16 9SS.}

\altaffiltext{4}{Physics Department, 
                 University of California, 
                 Santa Barbara, CA 93106.}

\altaffiltext{5}{Institute of Geophysics and Planetary Physics,
                 University of California,
                 Riverside, CA 92521.}

\altaffiltext{6}{Department of Physics and Astronomy,
                 University of California,
                 Riverside, CA 92521.}

\altaffiltext{7}{Department of Astronomy, 
                 University of California,
                 Berkeley, CA 94720-3411.}

\altaffiltext{8}{Lick Observatory,
                 P.O. Box 85, 
                 Mount Hamilton, CA 95140.}

\altaffiltext{9}{Department of Physics and Astronomy, 
                 University of California, 
                 Los Angeles, CA 90024.}

\altaffiltext{10}{Institute of Astronomy, 
                 School of Science, University of Tokyo, 
                 2-21-1 Osawa, Mitaka, Tokyo 181-0015, Japan.}

\altaffiltext{11}{Astronomy Program, Department of Physics and
                  Astronomy, Seoul National University, Gwanak-gu,
                  Seoul 151-742, Korea.}

\begin{abstract}
We present velocity-delay maps for optical \ion{H}{1}, \ion{He}{1},
and \ion{He}{2} recombination lines in Arp\,151, recovered by fitting
a reverberation model to spectrophotometric monitoring data using the
maximum-entropy method.  \ion{H}{1} response is detected over the
range 0--15~days, with the response confined within the virial
envelope.  The Balmer-line maps have similar morphologies but exhibit
radial stratification, with progressively longer delays for H$\gamma$
to H$\beta$ to H$\alpha$.  The \ion{He}{1} and \ion{He}{2} response is
confined within 1--2 days.  There is a deficit of prompt response in
the Balmer-line cores but strong prompt response in the red wings.
Comparison with simple models identifies two classes that reproduce
these features: freefalling gas, and a half-illuminated disk with a
hotspot at small radius on the receding lune.  Symmetrically
illuminated models with gas orbiting in an inclined disk or an
isotropic distribution of randomly inclined circular orbits can
reproduce the virial structure but not the observed asymmetry.  Radial
outflows are also largely ruled out by the observed asymmetry.  A
warped-disk geometry provides a physically plausible mechanism for the
asymmetric illumination and hotspot features.  Simple estimates show
that a disk in the broad-line region of Arp\,151 could be unstable to
warping induced by radiation pressure. Our results demonstrate the
potential power of detailed modeling combined with monitoring
campaigns at higher cadence to characterize the gas kinematics and
physical processes that give rise to the broad emission lines in
active galactic nuclei.
\end{abstract}


\section{Introduction}

Active galactic nuclei (AGNs) have long been known to undergo dramatic
and rapid variations in brightness \citep{matthews63,smith63}.  This
variability is exploited with the technique of reverberation
mapping \citep{blandford82,peterson93} to study the broad-line region
(BLR) of AGNs, a region of gas only $\sim 0.01$\,pc in size and
presumed to be photoionized by the emission from an accretion disk
around a supermassive black hole.  Spatially unresolvable even for the
most nearby galaxies, the BLR is studied in the time domain through
monitoring the variable continuum emission (presumably originating
close to the black hole) and its reprocessing in the BLR gas,
discernible as variations in the broad emission-line fluxes and
profiles.

Under the assumptions that (1) the continuum-emission region is much
smaller than the BLR, (2) time delays arise predominantly from
light-travel time, and (3) the relationship between the ionizing and
observed continuum is smooth, a linearized echo model can be used to
express changes in the emission lines ($\Delta L$) as time-delayed
responses to changes in the continuum ($\Delta C$):
\begin{equation} 
\label{eqn:echo}
\Delta L(v,t) \, = \, 
\int_{0}^{\infty} \Psi(v,\tau) \, \Delta C(t-\tau) \, d\tau
\ .
\end{equation}

\noindent Here, $\Delta L(v,t)$ is the change in the emission-line
flux at line-of-sight velocity $v$ and time $t$, $\Delta C(t)$ is the
change in the continuum flux at time $t$, and $\Psi(v,\tau)$ is the
so-called transfer function, or velocity-delay map, giving the
distribution of the line response over line-of-sight velocity $v$ and
time delay $\tau$ (i.e., the emission-line response to a
$\delta$-function continuum event).  Information on the BLR geometry,
kinematics, and ionization structure is encoded in $\Psi(v,\tau)$,
making recovery of velocity-delay maps a key goal of echo-mapping
experiments.

Due to the formidable requirements of long-duration monitoring, high
temporal sampling, and high-quality and homogeneous data for
recovery of velocity-delay maps \citep{horne04}, most echo-mapping
experiments have restricted the analysis to a simpler problem:
measuring the mean delay $\tau$, and hence the mean radius $R\approx
c\,\tau$ of the BLR, through cross-correlation of the line and
continuum light curves (e.g.,
\citealt{antonucci83,peterson83,kaspi00,peterson04}).  Analyses
comparing delays in different parts of the emission-line profiles,
such as the red wing versus the blue wing (e.g.,
\citealt{gaskell88,koratkar89,crenshaw90}), have sometimes shown
evidence of asymmetry, hinting at gross characteristics (inflow vs.
outflow vs. circulation) of the BLR gas flow.

For certain high-quality data sets, delay maps ($\Psi(\tau)=\int
\Psi(v,\tau)\, dv$) have been recovered.  These include delay
maps for Ly$\alpha$ and \ion{C}{4}, among other lines, derived from
the {\it International Ultraviolet Explorer} ({\it IUE}) monitoring
dataset for NGC\,5548 \citep{krolik91}, and the delay maps recovered
by \citet{horne91} from ground-based optical monitoring of H$\beta$ in
NGC\,5548.  The delay maps generally agree with the results of
cross-correlation analysis, while allowing additional insights into
the details of the BLR.
\citet{horne91} find a lack of H$\beta$ response at zero time delay in
NGC\,5548, a signature of anisotropic (inward) H$\beta$ response
and/or a paucity of BLR gas near the line of sight.

In addition to presenting delay maps for several ultraviolet lines in
NGC\,4151, \citet{ulrich96} also present a partially recovered
velocity-delay map for \ion{C}{4}, which unfortunately suffered from
strong intrinsic absorption in the line core throughout the campaign.
Three separate groups attempted to recover velocity-delay maps from an
{\it IUE} and {\it Hubble Space Telescope} monitoring campaign of
NGC\,5548; however, these studies were again hampered by the quality
of the \ion{C}{4} data used, and the final conclusions ranged from no
radial motion \citep{wanders95} to some radial infall \citep{done96}
to radial outflow (\citealt{chiang96,bottorff97}).

Recent results from the Lick AGN Monitoring Project (LAMP;
\citealt{bentz08,bentz09c}) and from MDM Observatory
(\citealt{denney09a,denney09b}) have shown that recovery of
velocity-delay maps may be within the reach of current AGN monitoring
programs due to their monitoring baselines of several months, daily
sampling, and homogeneous data with high signal-to-noise ratios
(S/N).  In this {\it Letter}, we present velocity-delay maps
recovered with maximum-entropy techniques for six optical H and He
recombination lines in the LAMP spectra of Arp\,151 (Mrk\,40):
H$\alpha$, H$\beta$, H$\gamma$, \heia, \heib, and \heii.  We also
suggest plausible models for the geometry and kinematics of the BLR in
Arp\,151 that could account for the observed velocity-delay structure.

\section{Observations}

Details of the photometric light curves are presented by
\citet{walsh09}.  In brief, broad-band Johnson $B$ and $V$ images of
Arp\,151 were obtained at the 0.8-m Tenagra II telescope in southern
Arizona between 2008 February 27 and May 16 (UT here and throughout).
Standard reduction techniques were employed, and photometric light
curves were measured with differential photometry relative to field
stars.  Absolute flux calibrations were determined using
\citet{landolt92} standard-star fields.

Spectroscopic monitoring of Arp\,151 was carried out at the Lick
Observatory 3-m Shane telescope with the Kast dual spectrograph
between 2008 March 25 and May 21 (see \citealt{bentz09c} for details).
The Kast red-side CCD with the 600~lines~mm$^{-1}$ grating allowed
spectral coverage over the range 4300--7100\,\AA.  Spectra were
obtained at a fixed position angle of 90\degr\ through a 4\arcsec-wide
slit, with a typical S/N $\approx 100$ per pixel at rest-frame
5100\,\AA.  Flux calibrations were determined from nightly
spectra of standard stars.

Small time-variable and wavelength-dependent corrections were made to
the spectra to account for three types of systematic error: (1)
wavelength shifts due to flexure and atmospheric differential
refraction, (2) spectral blurring due to atmospheric seeing and
instrumental resolution, and (3) photometric errors due to slit losses
and atmospheric transmission (see Bentz et al.\ 2010, in preparation,
for a detailed description of the methodology).
After correcting for these errors, the continuum-subtracted 
line-profile variations were subjected to echo-mapping analysis with
MEMECHO.

\section{MEMECHO Light-Curve Modeling}

MEMECHO (see \citealt{horne91,horne94} for details) uses a
maximum-entropy technique to recover $\Psi(v,\tau)$ by fitting the
linearized echo model of Equation~(\ref{eqn:echo}) to the observed
continuum light curve and responding emission-line velocity
profiles. MEMECHO finds the ``simplest'' positive image $p_i$ that
fits the data. The image $p_i$ includes an evenly sampled driving
continuum light curve $C(t)$, a background spectrum $L_0(\lambda)$
representing nonvariable line emission, and the velocity-delay map
$\Psi(v,\tau)$. The quality of the fit to the $N$ data points is
measured by $\chi^2$ summed over measurements of the driving continuum
light curve and the reverberating emission-line spectrum. Simplicity
is measured by the entropy $S=\sum_i p_i - q_i - p_i \ln(p_i/q_i)$
relative to a ``default image'' $q_i$.  Since $\partial S/\partial p_i
= - \sum_i \ln(p_i/q_i)$, maximum entropy at $S=0$ occurs when
$p_i=q_i$.  Our default image $q_i$ slightly blurs the image $p_i$
(for example $q_i=\sqrt{p_{i-1}\,p_{i+1}}$ for a one-dimensional
image), and thus $S$ increases toward 0 for smooth positive images.
With suitable datasets MEMECHO delivers a well-defined map maximizing
$S$ for a specified $\chi^2/N$.  The resolution of the map and the
quality of the fit improve as $\chi^2/N$ is reduced.  When $\chi^2$ is
set too low, the fit attempts to follow noise in the data, resulting
in noisy maps.  The maps we show are for $\chi^2/N=1.1$, giving the
highest resolution maps and the best fits we could achieve while
avoiding the introduction of spurious structure, typically large
excursions in the gaps between measurements of $C(t)$.

We chose the $B$-band photometric light curve as the driving continuum
light curve for the MEMECHO analysis. The variations in $B$ are
stronger than in $V$ with no evident time delay between the two
\citep{walsh09}, and the photometric light curves are more accurately
calibrated than the spectroscopic continuum light curve, which suffers
from aperture effects.  Furthermore, the contribution of the broad
emission lines to the photometric variability is negligible (see
\citeauthor{walsh09} for a complete discussion).  The $V$-band light
curve and the integrated H$\alpha$, H$\beta$, and $H\gamma$ light
curves were then modeled as echoes of $B$, as were the emission-line
light curves at each wavelength.  Thus, we fit delay distributions
simultaneously to a total of 1373 individual light curves --- one per
spectral pixel (each with $\Delta\lambda = 2$\,\AA), plus the $V$-band
and integrated Balmer-line light curves.  We enforce causality by
requiring non-negative delays, $\tau_{\rm min} = 0$\,days.  We set
$\tau_{\rm max} = 20$\,days, which is $\sim 1/3$ of the spectroscopic
monitoring baseline, and we use a uniform spacing of $\Delta t =
0.5$\,day in the delay maps and continuum light curve.  The linearized
echo model gives a tangent-line approximation to what may in principle
be nonlinear responses to changes $\Delta C(t) = C(t)-C_0$ away from a
reference continuum level $C_0$.  For $C_0$ we adopt the mean of the
observed $B$-band fluxes.

In Figure~1, we show selected light curves of Arp\,151 along with
their MEMECHO fits and recovered delay maps.  The selected wavelengths
sample the cores of the emission lines plus the Balmer-line wings.
The $B$-band (driving) light curve is shown in the bottom panel.  The
value of $\chi^2/N$ is reported for each light curve shown, and is 1.1
when summed over all light curves including those not shown.

\section{The Velocity-Delay Maps}

The delay maps at each wavelength are assembled into a velocity-delay
map shown over the full spectral range in Figure~2 and centered on the
six strongest emission lines in Figure~3.  To aid in comparing the
velocity-delay structure of different emission lines, Figure~4 presents
false-color maps with red, green, and blue colors giving the
velocity-delay distributions of different emission lines as indicated.

Taken together, the velocity-delay maps exhibit a virial structure
with the emission-line response covering a wider range of velocities
at smaller delays. This is perhaps best seen in Figures~3 and 4, where
the line responses are confined within the ``virial envelope''
$v^2=G\,M/c\,\tau$ for $M_{\rm virial}=1.2\times10^6\,{\rm M}_\odot$
\citep{bentz09c}.

The Balmer-line maps all have the same basic structure, with longer
delays in the core and smaller delays in the line wings, and with the
red-wing response much stronger than that of the blue wing.  The
H$\alpha$ response is detected from 0 days in the red wing to about 15
days in the line core. The other Balmer lines have a similar
structure, but with H$\beta$ extending to about 7 days and H$\gamma$
to 5 days.  The mean delays from these maps are consistent with the
mean lag times from the cross-correlation analysis presented by
\citet{bentz10}.  A plausible interpretation of this effect is a
radial stratification resulting from optical-depth effects within the
Balmer series \citep{korista04}.

The response of the helium lines is consistent with zero time delay,
except for some response at small positive delays in the
map of \heia\ and \heib.  As discussed by \citet{bentz10}, the helium
response occurs on shorter timescales than our monitoring sampling and
is therefore unresolved.

The H$\alpha$ map shows a ``curl'' at long time delays on the blue
side of the line outside the virial envelope.  This weak feature may
be an artifact, as it does not always appear in maps made with slightly
different fitting parameters.  Features that appear in all of the
Balmer lines, such as the strong emission in the red wings, are more
secure.

The maps show a lack of prompt response in the core of all three
Balmer lines.  The delay map for the integrated H$\beta$ profile in
Figure~1 shows a paucity of response at zero lag, similar to what was
found for NGC\,5548 \citep{horne91}, even though the full
velocity-delay map clearly shows gas responding with zero lag in the
red wing of H$\beta$.

\section{Discussion}

We have examined the predicted velocity-delay structure for a variety
of simple BLR models to identify classes of models that fail and
succeed in qualitatively reproducing the main features of the observed
velocity-delay maps. Two successful classes of models are represented
by the Freefall and Disk~+~Hotspot models shown in Figure~5.  For all
four models in Figure~5 we adopt $M_{\rm BH}=7\times10^6~{\rm M}_\odot$
(assuming $f=5.5$; \citealt{bentz09c}).

A thin spherical shell of radius $R$ infalling with velocity $V$
covers a sloped line on the velocity-delay map, with
$\tau=(1-\cos{\theta})R/c$ and $v=V\cos{\theta}$, the delay increasing
from $\tau=0$ at $v=+V$ on the near side to $\tau=2\,R/c$ at $v=-V$ on
the far side of the shell.  The Freefall model in Figure~5 has a
spherical distribution of infalling gas with $R_{\rm in} =
0.1$~lt-days and $R_{\rm out} = 4$~lt-days, and $V=\sqrt{GM/R}$,
giving small delays on the red side and longer delays with a virial
envelope on the blue side of the map. This naturally produces the
required asymmetry. The lack of prompt response in the line core can
arise by means of an inward radiation anisotropy (as displayed in
Figure~5 with $F_{\rm in}/F_{\rm total}=0.8$) and/or by reducing the
response of the inner gas.  Models with radially outflowing gas have
the opposite asymmetry and are thus largely ruled out.

For an inclined circular orbit the velocity-delay structure is an
ellipse with $\tau=(1+\sin{i}\cos{\theta})R/c$ and
$v=V\sin{i}\sin{\theta}$.  The basic virial envelope, with a wider
velocity range at smaller delays, can be reproduced by distributions
of circular Keplerian orbits ranging from isotropically distributed
angular momentum to inclined disks (e.g., the Thick Spherical Shell
and Isotropically Illuminated Disk models in Figure~5). For such models
the Balmer response at small delay in the line core can be reduced by
(1) anisotropic ionizing radiation directed toward the far side, (2)
anisotropic inwardly directed line response, and/or (3) excluding
edge-on orbits to form an inclined disk-like geometry.  The enhanced
prompt response in the red wing is hard to reproduce with a symmetric
distribution of circular orbits.  If the orbits circulate in the same
direction, however, then the required asymmetry can be introduced by
azimuthal structure-enhancing response (a ``hot spot'') on the
receding side of the inner disk. While this is somewhat {\it ad hoc},
it might plausibly arise in the context of a warped-disk geometry, the
warp exposing gas to ionizing radiation at small radius while
shielding it at larger radius.  The Disk~+~Hotspot model displayed in
Figure~5 is inclined ($i = 20\degr$) with $R_{\rm in} = 0.1$~lt-days and
$R_{\rm out} = 4$~lt-days.  No emission originates from the near side
of the disk due to its warped shape, and excess emission arises on the
receding portion of the inner disk at 0.1--0.3~lt-days.

Both the Infall model and Disk~+~Hotspot model qualitatively reproduce
the observed features of the velocity-delay maps, and more work is
necessary to determine which family of models is preferred.  In the
meantime, we note that
warped disks are seen in many astrophysical environments 
and have previously been invoked to explain various AGN phenomena
(e.g., flux variations in the double-peaked broad emission from
NGC\,1097, \citealt{storchibergmann97}; Fe emission-line profiles in
the X-rays, \citealt{hartnoll00}; the misalignment of radio jets and
nuclear disks, \citealt{schmitt02}; and a self-contained apparatus for
AGN unification, e.g., \citealt{nayakshin05}).  We consider here the
plausibility of irradiation-induced warping \citep{pringle96}, in
which an optically thick, geometrically thin disk irradiated by a
central source is unstable to warping from radiation pressure.  The
instability criterion is
\begin{equation}
\frac{R}{R_G} \, \geq \, 8 \pi^2 \, \eta^2 \, \epsilon^{-1} ,
\end{equation}
\noindent where $R_G \, = \, GM_{\rm BH}/c^2$, $\eta$ is the ratio of
the viscosities in the plane of and perpendicular to the disk (assumed
here to be equal, i.e., $\eta=1$), and $\epsilon$ is the accretion
efficiency.  Assuming $\epsilon \, \approx \, 0.1$, a disk in the BLR
of Arp\,151 could become unstable on length scales $R/R_G \, \geq \,
800 \, \approx 0.3$\,lt-day, inside the scale of the optical BLR.  For
typical estimates of BLR parameters (following
\citealt{storchibergmann97}), the precession timescale would likely be
decades to centuries.


\section{Summary}

We have recovered velocity-delay maps for six optical hydrogen and
helium recombination lines in the LAMP spectra of Arp\,151 using the
maximum-entropy method. These are the most detailed velocity-delay
maps constructed to date.  The individual Balmer lines have similar
velocity-delay structure, and the maps show responses with delays
increasing from H$\gamma$ to H$\beta$ to H$\alpha$.  This can be
interpreted as radial stratification through the Balmer series and is
expected from optical-depth effects.  All three lines show prompt
response in their red wings, contrary to what has been seen in
NGC\,5548.  The helium-line response is mostly unresolved in time.

The features seen in the Balmer-line velocity-delay maps can be
reproduced qualitatively with either of two simple {\it ad hoc}
models: a freefalling BLR, or a partially illuminated thin disk with a
localized excess of emission at small radii that could be interpreted
as a warped disk.  Warped disks are known to occur in many
environments and on many astronomical scales, and it is plausible that
the BLR in AGNs is another manifestation of this common configuration,
as has been previously suggested in the literature.  It appears to be
impossible for a simple disk model or a thick spherical shell composed
of randomly inclined circular orbits to accurately reproduce the
features seen in the velocity-delay maps, including the strong
emission at short lag times in the red wings of the Balmer lines and
the lack of emission at short lag times in the line cores. Radial
outflows are also largely ruled out by the observed asymmetric
response.

The simple models and plausibility arguments presented here are the
first steps in interpreting the velocity-delay maps for Arp\,151.  We
are currently investigating other LAMP targets and expect to recover
velocity-delay maps for a few additional objects, which may lead to
insights into BLR differences across the Seyfert~1 population.  Future
monitoring programs will benefit from even higher temporal sampling
and the avoidance of any gaps in the driving continuum light curve,
allowing the driving light curve to provide stronger constraints and
leading to more detailed velocity-delay maps of the emission-line
responses.  More sophisticated modeling will also help with
interpreting the velocity-delay maps and determining which families of
BLR models can be ruled out, allowing additional insights into the
complexities of AGN BLRs.

\acknowledgements

We are grateful to Brad Peterson, Kelly Denney, and Kirk Korista for
helpful conversations.  M.C.B.\ thanks the University of St.\ Andrews
for their hospitality and the STFC Rolling Grant to St.\ Andrews for
supporting a visit that enabled the beginnings of this investigation.
M.C.B.\ gratefully acknowledges support provided by NASA through
Hubble Fellowship grant HF--51251 awarded by the Space Telescope
Science Institute, which is operated by the Association of
Universities for Research in Astronomy, Inc., for NASA, under contract
NAS 5-26555.  LAMP was supported by NSF grants AST--0548198 (UC
Irvine), AST--0607485 and AST--0908886 (UC Berkeley), AST--0642621 (UC
Santa Barbara), and AST--0507450 (UC Riverside).


\begin{figure*}
\epsscale{0.85}
\plotone{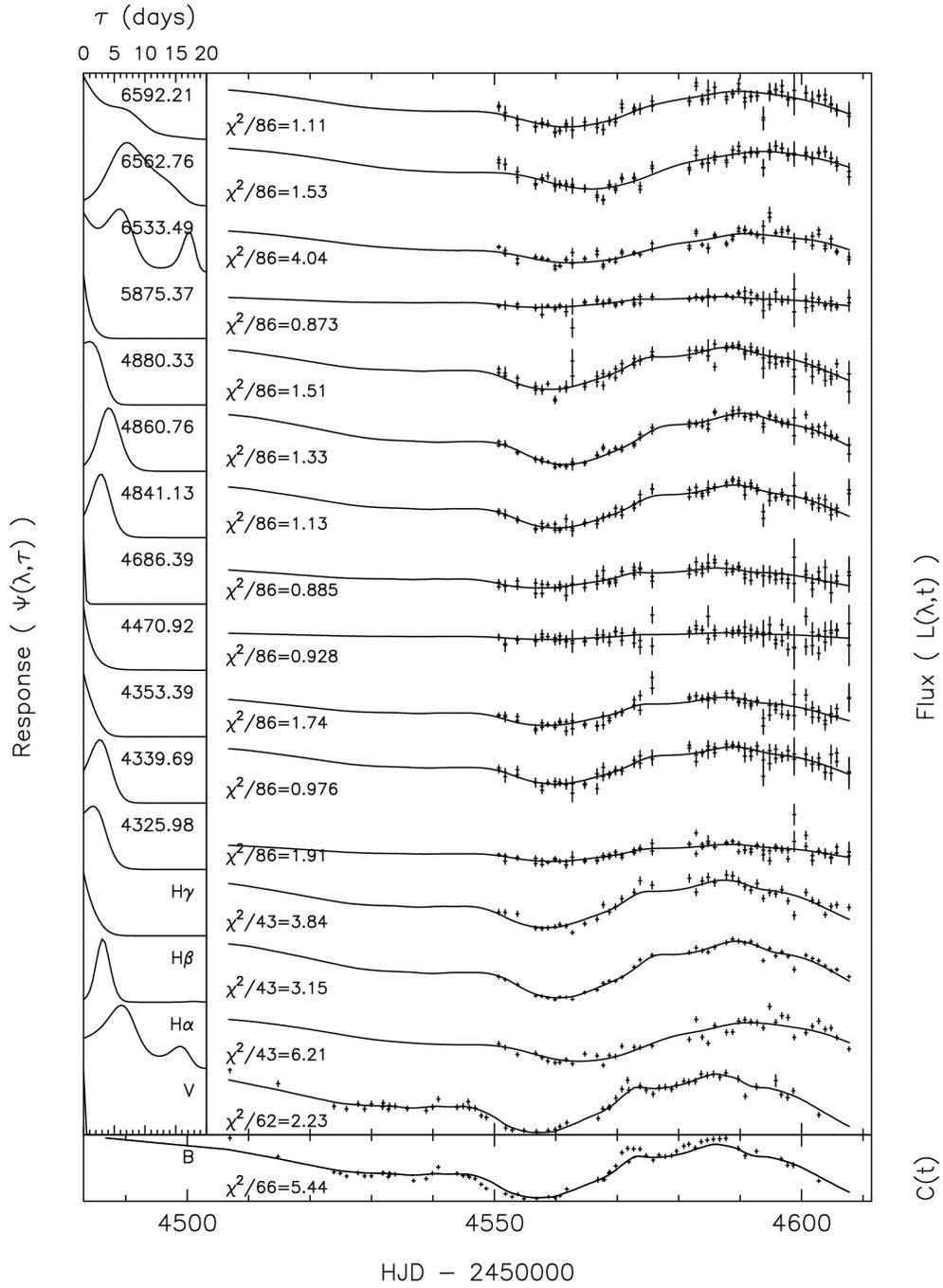}
\caption{MEMECHO fits to the light curves at selected wavelengths in
  the spectra of Arp\,151 ({\it right panel}) and their delay maps
  ({\it left panel}).  The rest-frame central wavelength of each
  pixel is listed with its delay map, and the $\chi^2/N$ values are 
  listed with the light curves.  We include four supplementary
  integrated light curves ($V$ band, H$\alpha$, H$\beta$, and H$\gamma$)
  as fitting constraints, which are displayed near the bottom.
  Multiple measurements in the same night are averaged together in the
  integrated light curves, but are treated separately for the
  non-integrated light curves.}
\end{figure*}

\begin{figure*}
\epsscale{1}
\plotone{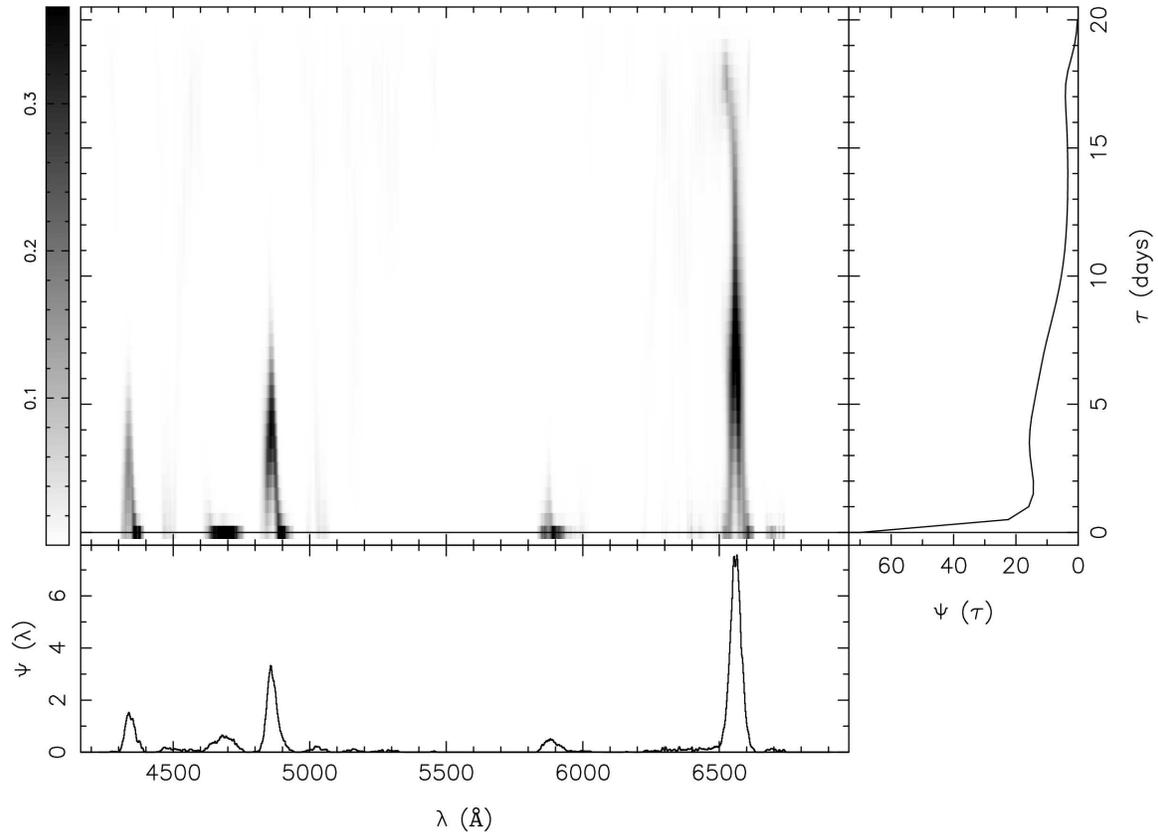}
\caption{Velocity-delay map $\Psi(\lambda,\tau)$ over the full
  wavelength range {\it (greyscale)}, and projections onto
  the wavelength axis $\Psi(\lambda)=\int\Psi(\lambda,\tau)\,d\tau$
  {\it (bottom)} and the time-delay axis
  $\Psi(\tau)=\int\Psi(\lambda,\tau)\,d\lambda$ {\it (right)}.}
\end{figure*}

\begin{figure*}
\plotone{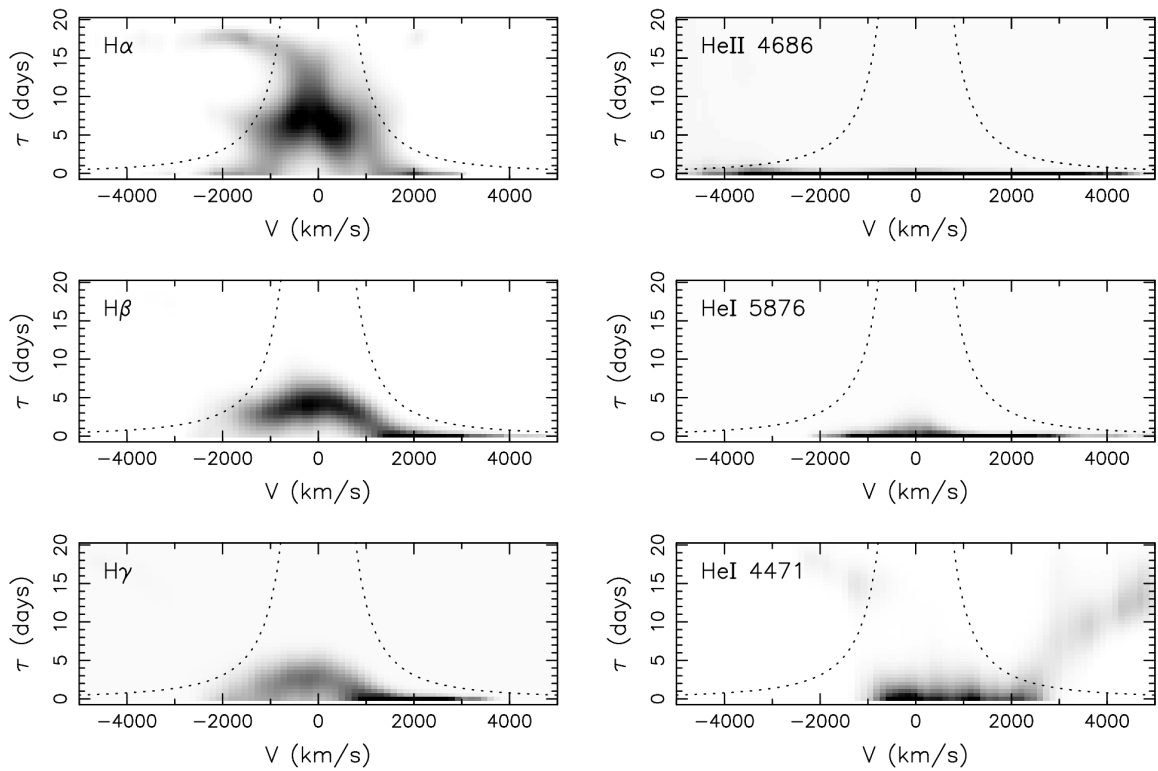}
\caption{Velocity-delay maps for each of the six optical H and He
  recombination lines in the LAMP spectra of Arp\,151. The dotted
  lines in each panel show the ``virial envelope'' $V^2\tau\,c/G= 1.2
  \times 10^6 {\rm M}_{\odot}$, based on the ``virial product'' of time lag
  and line width for H$\beta$ \citep{bentz09c}.}
\end{figure*}

\begin{figure*}
\epsscale{1}
\plottwo{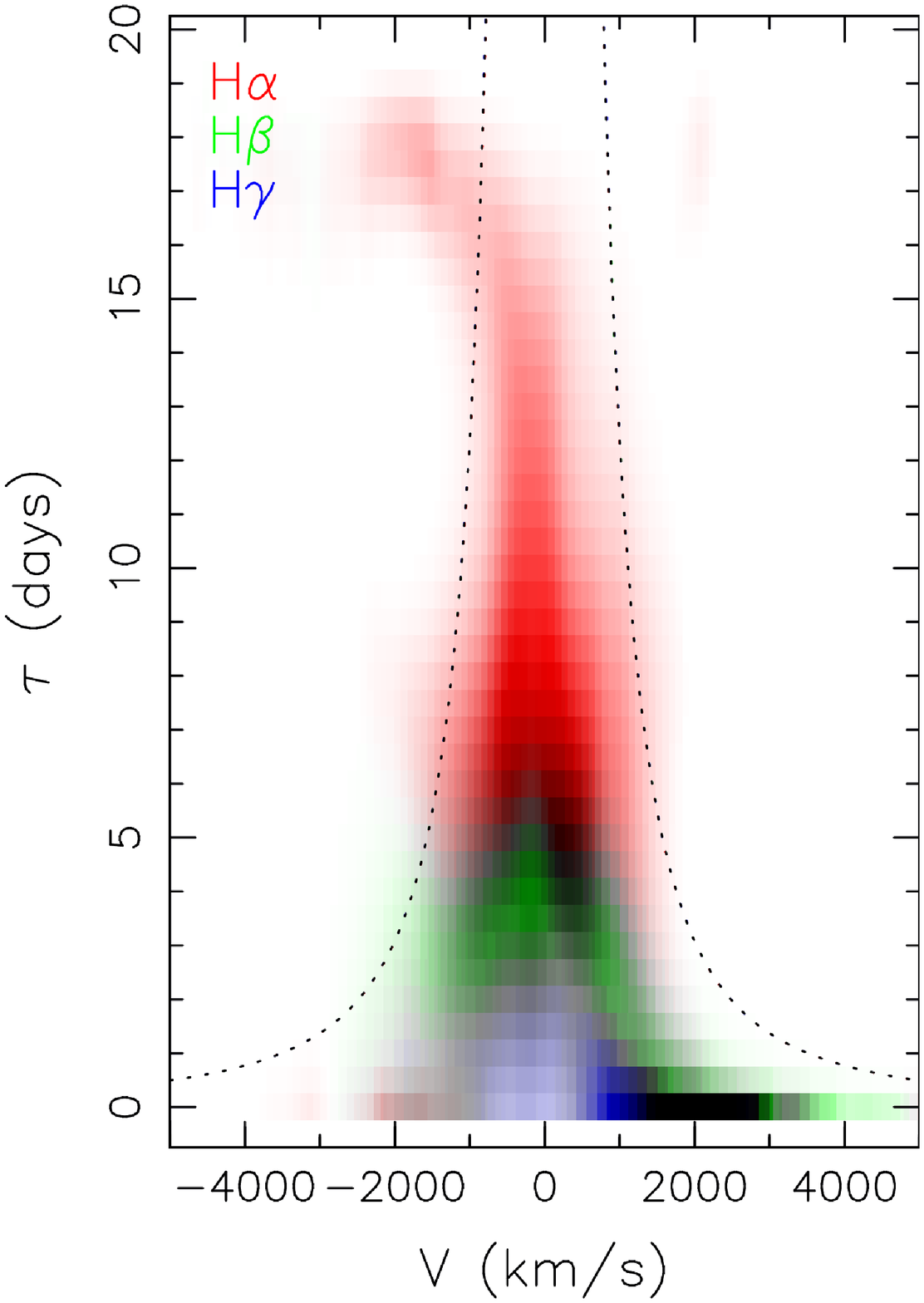}{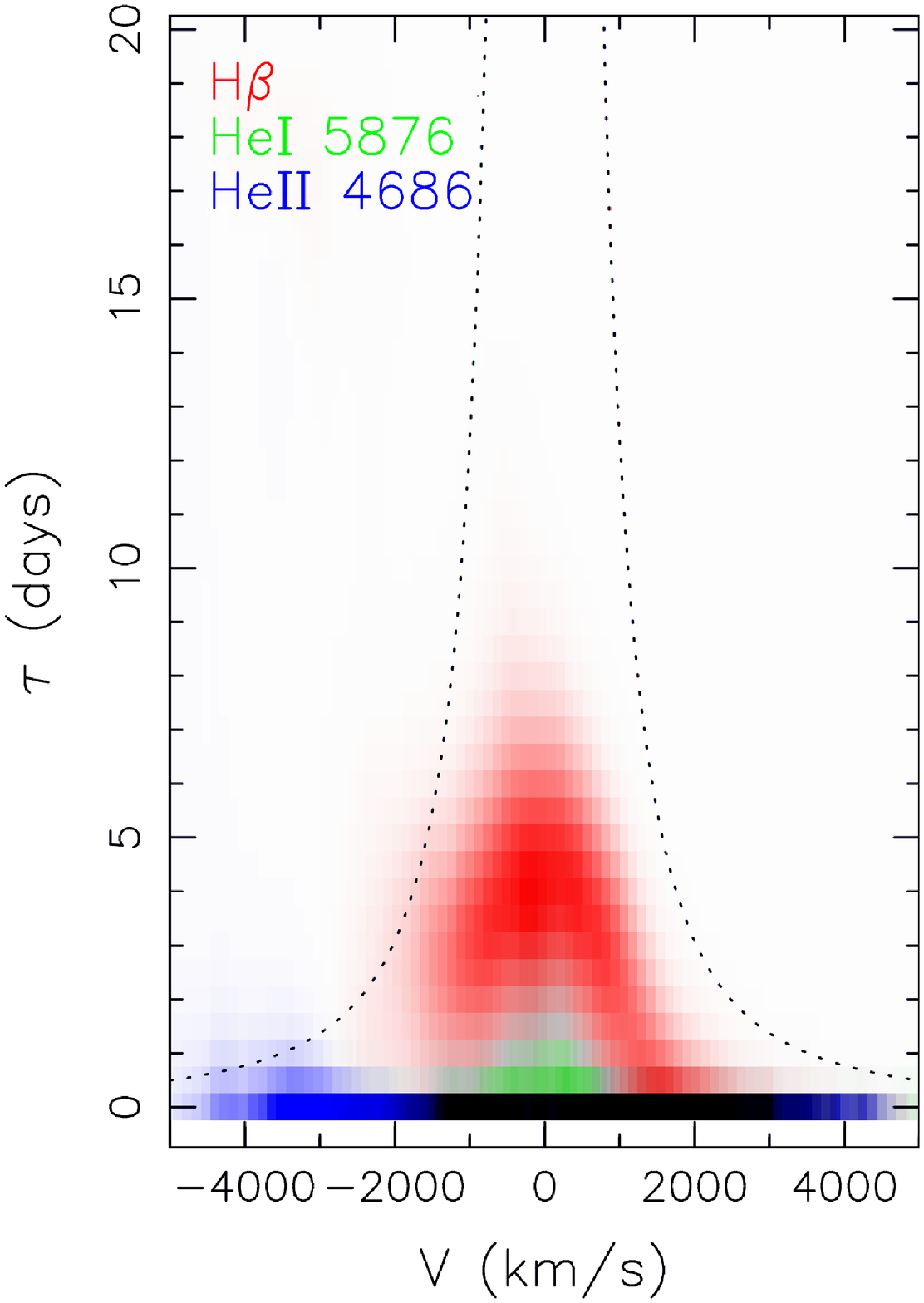}
\caption{{\it Left:} Comparison of the velocity-delay maps for
  H$\alpha$ (red), H$\beta$ (green), and H$\gamma$ (blue).  {\it
    Right:} Comparison of the velocity-delay maps for H$\beta$ (red),
  \ion{He}{1} $\lambda 5876$ (green), and \ion{He}{2} $\lambda 4686$
  (blue).  As in Figure~3, the dotted lines in each panel show the
  virial envelope.  The Balmer lines all have the same basic shape,
  but are graduated in their response ranges.  The helium-line
  response is mostly unresolved in time.}
\end{figure*}

\begin{figure*}
\plotone{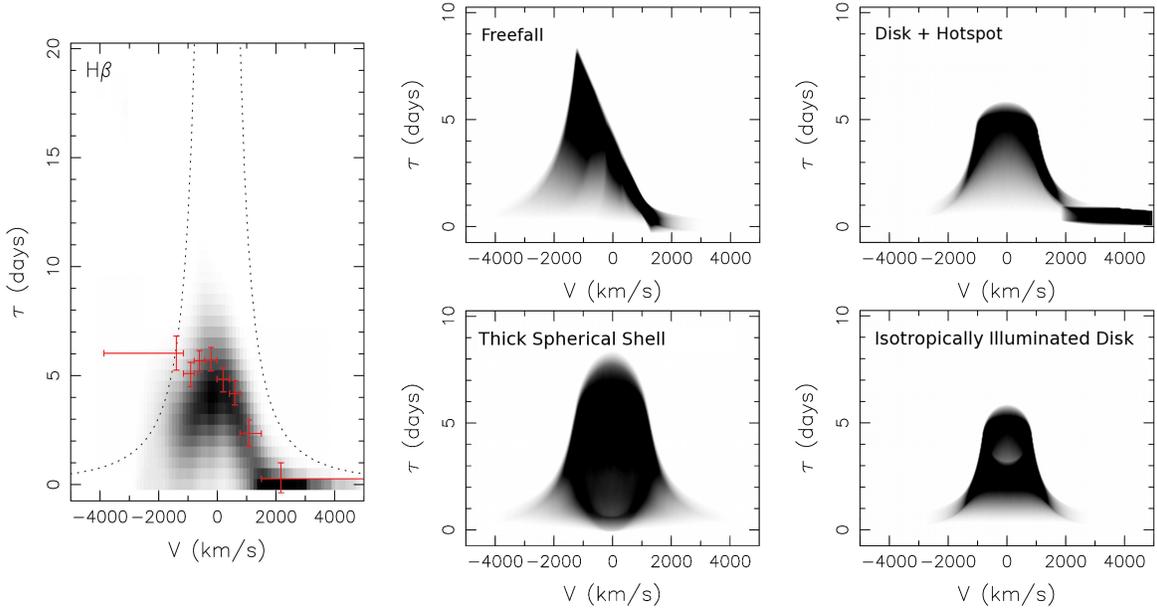}
\caption{Velocity-delay map for H$\beta$ compared to simple models for
  $M_{\rm BH}=7\times10^6 {\rm M}_{\odot}$: {\it (top)} a freefalling BLR
  with inward emission anisotropy {\it (left)}, and a partially
  illuminated disk with a hot spot at small radii on the receding side
  {\it (right)}; {\it (bottom)} a biconically illuminated thick shell
  with circular orbits in a Keplerian potential and inward emission
  anisotropy {\it (left)}, and a fully illuminated disk with inward
  emission anisotropy {\it (right)}.  The top models more accurately
  reproduce the features seen in the H$\beta$ velocity-delay map than
  do the bottom models.  All models have been smoothed
  to match the lower resolution of the recovered velocity-delay maps.
  The dotted lines in the H$\beta$ panel again show the virial
  envelope, and the red data points show the average lag measurement
  per velocity bin from \citet{bentz09c}.}
\end{figure*}

\end{document}